\documentclass[11pt,twoside]{article}
\usepackage{asp2006}
\usepackage{epsf}
\usepackage{graphics}
\usepackage{lscape}
\usepackage{graphicx}
\def\edcomment#1{\iffalse\marginpar{\raggedright\sl#1\/}\else\relax\fi}
\reversemarginpar

\begin{document}
\title{The radio properties of type II quasars}
\author{Dharam Vir Lal$^1$ and Luis C. Ho$^2$}
\affil{$^1$Institute of Astronomy \& Astrophysics, Academia Sinica,
Taipei 10617. \\
$^2$The Observatories of the Carnegie Institution of Washington, 813 Santa Barbara St., Pasadena, CA 91101.}

\begin{abstract}
Quasars (of type I) are the luminous analogs of type I Seyfert galaxies.  
Within the framework of unified models of active galaxies, the population
of quasars of type II recently discovered with the Sloan Digital Sky Survey
are the luminous analogs of type II Seyfert galaxies.
Since our knowledge and understanding of the radio properties of
these type II quasars are very limited, we have performed Very Large Array
observations for a sub-sample of such sources.
Our detection rate of 61\% is consistent with
the detection rate for other AGN samples.
We do not find a correlation between radio and
[O~III]~$\lambda$5007 luminosities for these sources.
Although the distribution of spectral indices is similar to that of the
3C sources, the lack of dependence of radio luminosity 
on [O~III]~$\lambda$5007 luminosity suggests
that not all sources in the sample are genuine AGNs.
\end{abstract}
\thispagestyle{plain}

\section{Introduction}

\vspace*{-0.25cm}
The standard model for radio-emitting
active galactic nuclei (AGNs; Scheuer 1974) suggests that there is a
supermassive black hole located in the center of the galaxy and a pair
of relativistic (most probably) electron-positron continuous beams, or jets,
expanding into the interstellar and intergalactic medium of the source (Begelman
et al. 1984).  Our current understanding of many of the 
collective properties of
AGNs can be summarized by so-called unification models
(e.g., Antonucci 1993; Urry \& Padovani 1995), in which the observed properties
of active galaxies are governed primarily by orientation and intrinsic
luminosity.  Many of the apparent differences between type I (broad-line) and
type II (narrow-line) AGNs are believed to be due to our line-of-sight having
different orientations with respect to the disk.
In unified models of active galaxies, AGN nuclei are surrounded by a dusty
molecular torus. In type II AGNs, with the view being edge-on, the torus
blocks a direct view of the continuum and broad-line region, which can only be
detected
through light scattered into the line-of-sight by material lying directly above
the torus opening; by contrast, in type I AGNs, the view being pole-on, the
continuum and broad-line region can be viewed directly.
This is indeed the case for most Seyfert galaxies (Lal et~al. 2004), and, in a
similar fashion, the same unified model must apply to higher-luminosity
AGNs such as quasars. Therefore, there should exist high-luminosity obscured
AGNs (type II quasars), which would be observable up to high redshifts.  These
also have been postulated for several other reasons, for instance 
to account for the cosmic hard X-ray background
(see Madau, et~al. 1994; Zdziarski et al. 1995).

\section{Our goal}

\vspace*{-0.25cm}
The Sloan Digital Sky Survey (SDSS; York et al. 2000) makes it possible
to find a large number of type II quasar candidates, and
now, following Zakamska et~al. (2003) we have a large sample
(291 objects) of known/candidate type II quasars at redshifts
0.3 $<$ $z$ $<$ 0.83.
Our understanding of the radio properties of type II quasars is based primarily
on FIRST survey results (Ivezi\'c et al. 2002; Zakamska et al. 2004), where
less than half of the sample sources have matches with the FIRST catalog. The
results showed no radio source with complex morphology, and overall only 17
sources (see typical maps in Fig. 1) have radio structure that appears extended.
This incomplete understanding of the radio properties of type II quasars
forms the motivation of this project, which aims to obtain 
high-angular resolution and high-sensitivity observations of these sources.
Given the cosmological importance of this class of AGNs and how little is
currently known about their basic properties, our primary goal
is to establish their observational radio properties.

\vspace*{-0.25cm}
\section{Sample of type II quasars and the observations}

\vspace*{-0.25cm}
Our sample is drawn from the type II quasars sample of Zakamska et~al.
(2003). This is the first complete sample of such sources, for which we
understand the optical properties in great detail.  Our control sample
of type I quasars is drawn from the Palomar-Green (PG) survey
(Schmidt \& Green 1983), whose radio
properties have already been well studied (Kellermann et al. 1989, 1994;
Miller et al. 1993).  The two samples have similar redshift range, and we
further match them in an orientation-independent parameter, namely
[O III]~$\lambda$5007 luminosity.

The radio observations were carried out in snapshot mode in a single observing
run on July 24--25, 2006, with the VLA in B configuration, with two 50 MHz
IFs at a mean frequency of 8.4351 GHz. This provided a typical resolution of
$\sim$ $0\hbox{$.\!\!^{\prime\prime}$}6$.
Apart from our new data, we also used the FIRST results at 1.4 GHz
to compare the results, and were able to gain VLA B configuration
images for 57 of the 59 sources at a typical resolution of
$\sim$ $6\hbox{$.\!\!^{\prime\prime}$}0$.

\vspace*{-0.25cm}
\section{Preliminary results}

\vspace*{-0.25cm}
Our results are shown in Figure 1.
The morphology of the radio emission is predominantly that of a compact core,
either unresolved or slightly resolved, occasionally accompanied by elongated,
jetlike features.

\begin{figure}[ht]
\begin{center}
\resizebox{12.5cm}{!}{\includegraphics{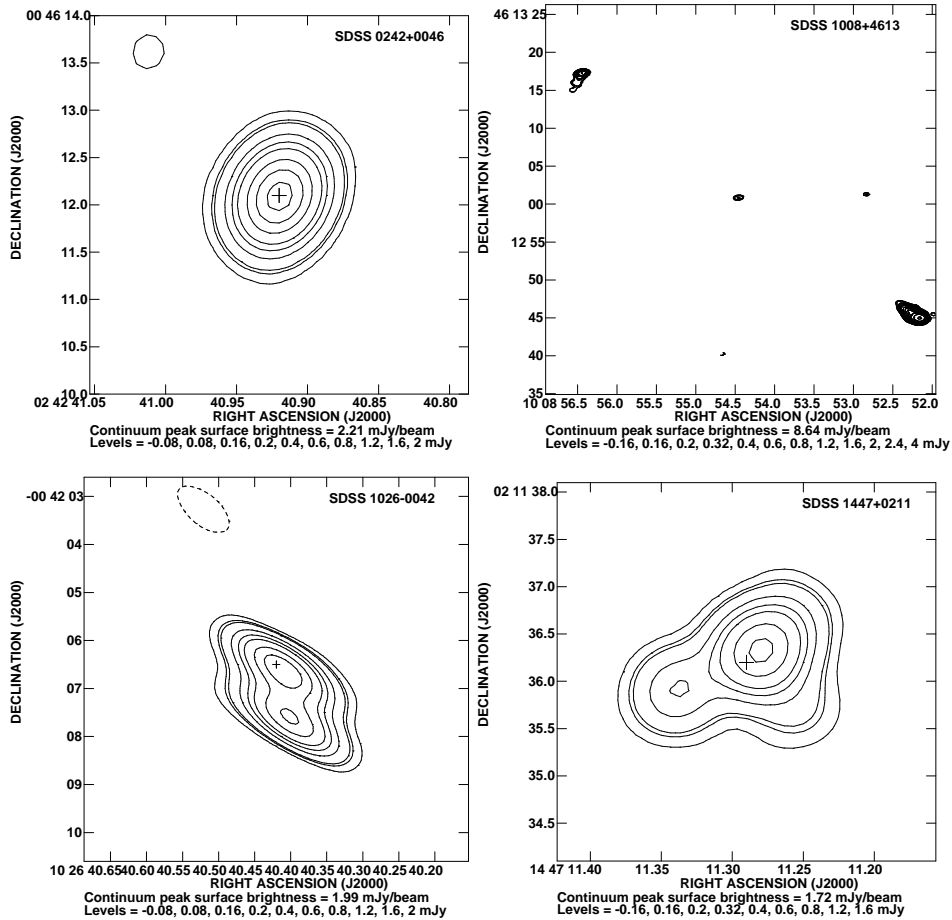}}
\end{center}
\caption{Radio morphology of four representative sources from our list
of 36 detected sources. The sample shows a range of radio morphologies: 
unresolved, partially resolved, diffuse or ambiguous,
and classical double radio sources. The fields are centered on the
positions given by Zakamska et~al. (2004), indicated by cross marks.
Image fields are 4 $\times$ 4 arcsec$^2$, except for SDSS~1008$+$4613,
which is 50 $\times$ 50 arcsec$^2$.}
\end{figure}

We detected 36 of the observed 59 sample sources.  This detection rate
(61\%) is slightly lower than, but not very different from,  the detection
rate for the PG sample (84\%), and can possibly be improved using the
technique of stacking (Glikman et~al. 2004, Greene et~al. 2006).

\begin{figure}[ht]
\begin{center}
\resizebox{13.2cm}{!}{\includegraphics{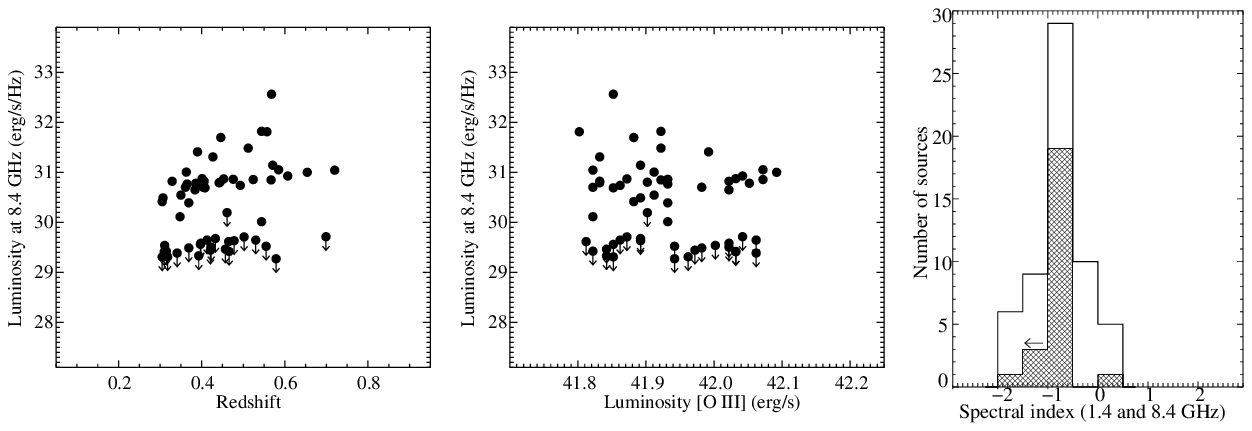}}
\end{center}
\caption{Figure showing source dependences and spectral index distribution.
Left: Schematic illustration of the coverage of the luminosity (at 8.4 GHz)
vs. redshift.
Middle: Plot of [O III]~$\lambda$5007 luminosity vs. radio luminosity
(at 8.4 GHz), showing absence of any strong correlation.
Right: The distribution of spectral index ($S_\nu \propto \nu^\alpha$,
between 8.4 and 1.4 GHz); the hashed region
indicates sources with non-detections.}
\end{figure}

The simplest sample type
has as its selection criteria only a single flux limit
in a chosen observing band and a chosen sky area. However, for any single
flux-limited sample chosen in this way there will be an inevitable and tight
correlation between luminosity and redshift.
Although we did not focus to fill the luminosity and redshift plane,
no visible correlation is seen in our sample (Fig. 2, left panel).
Or in other words, it is therefore possible to determine in our
sample whether the dependence of a given source property
is primarily on redshift or on luminosity (Blundell et~al. 1999).

Figure 2 (middle panel) shows the [O III]~$\lambda$5007 luminosities versus
the radio luminosities of the observed sample sources.
No obvious dependence of the radio luminosity of an object on
its [O III]~$\lambda$5007 luminosity is seen, although many objects show
upper limits on radio luminosities.
This is surprising because both radio and [O III]~$\lambda$5007 luminosities
are indicators of intrinsic AGN power, and Seyfert galaxies
predominantly show a strong correlation (Whittle 1992).
We therefore suspect that not all of the sample sources are dominated by 
AGN activity, in line with the suggestion by Kim et al. (2006) that these 
sources tend to show enhanced star formation.

In addition to full-resolution maps, we also make tapered maps
matched to the resolution of FIRST and use them to determine the integrated
spectral indices.
The distribution of spectral indices is shown in Figure 2 (right panel),
and the median of it is similar to that of the 3C radio sources.

\noindent
{\bf {Acknowledgments}}

\smallskip
\noindent
The VLA is operated by the US National Radio Astronomy Observatory which is
operated by Associated Universities, Inc., under cooperative agreement with
the National Science Foundation.

\end{document}